\documentstyle[aps,prd,twocolumn,epsf,epsfig]{revtex}
\newcommand \bea{\begin{eqnarray}}
\newcommand \eea{\end{eqnarray}}
\newcommand \beq{\begin{eqnarray}}
\newcommand \eeq{\end{eqnarray}}
\newcommand \ga{\raisebox{-.5ex}{$\stackrel{>}{\sim}$}}

\begin{document}
\twocolumn[\hsize\textwidth\columnwidth\hsize
\csname@twocolumnfalse%
\endcsname
\draft
\title{Cold Bose gases with large scattering lengths}
\author{S.\ Cowell$^1$, H.\ Heiselberg$^2$, I.\ E.\ Mazets$^3$, 
J.\ Morales$^1$, V.\ R.\ Pandharipande$^1$, and C.\ J.\ Pethick$^2$}
\address{
{\setlength{\baselineskip}{18pt}
$^1$\,Department of Physics, University of Illinois at Urbana-Champaign, 
1110 West Green Street, Urbana, IL 61801-3080,\\
$^2$\,NORDITA, Blegdamsvej 17, DK-2100 Copenhagen \O, Denmark,\\
$^3$\,Ioffe Physico-Technical Institute, 194021 St.Petersburg, Russia.}
}
\maketitle

\begin{abstract}
We calculate the energy and condensate fraction for a dense system of
bosons interacting through an attractive short range interaction with
positive s-wave scattering length $a$.  At high densities, $n\gg
a^{-3}$, the energy per particle, chemical potential, and square of
the sound speed are independent of the scattering length and
proportional to $n^{2/3}$, as in Fermi systems.
The condensate is quenched at densities $na^3\simeq 1$.\\
\end{abstract}
\vskip1pc]

The densities of atomic vapors under experimental
conditions are generally low, in the sense that the distance
between particles is typically much larger than the magnitude of the
atomic scattering length $a$, which characterizes the strength of
two-body interactions.  Recently, it has become possible to tune
atomic scattering lengths to essentially any value, positive or
negative, by exploiting Feshbach resonances\cite{feshbach}.  This opens
up the possibility of creating systems which are dense, in the sense
that the ``gaseousness parameter", $n|a|^3$, where $n$ is the particle
density, is large compared with unity \cite{depletion}.

In dilute alkali gases, the bare two-body interaction at distances
large compared with the atomic size is attractive due to the van der
Waals interaction, but the effective two-body interaction at low
energies, which is given by $4\pi\hbar^2a/m$, may be large and positive
due to the existence of a bound state just below threshold.

One of the novel features of such gases is that the
characteristic length $R$ associated with the range of the interatomic
potential is of order $10^{-6}$ cm.  The quantity $r_0=(3/4\pi
n)^{1/3}$, which is a measure of the atomic separation, is typically
of order $10^{-4}$ cm, and therefore much larger than $R$.  In this
Letter we consider the properties of a Bose gas for arbitrary
$n|a|^3$, assuming that $R\ll r_0$.  The corresponding problem for a
Fermi gas has been considered previously \cite{HH}.

We begin by reviewing briefly the properties of the dilute Bose gas
with repulsive interactions, $a>0$.
At low densities ($na^3\ll 1$) the interaction energy
is proportional to the scattering length and the density, as was first
found by Lenz \cite{Lenz}. With higher order terms calculated by Huang,
Lee, and Yang \cite{LY} and Wu \cite{Wu} included, the energy per particle is
\bea
   \frac{E}{N} &=& 2\pi \frac{\hbar^2a}{m} n 
   \left[1+\frac{128}{15}\left(\frac{na^3}{\pi}\right)^{1/2} \right. 
   \nonumber\\
   && \left. +8\left(\frac{4\pi}{3}-\sqrt{3}\right)
 na^3\ln(na^3)+{\cal O}(na^3) \right]\,. \label{dilute}
\eea
To this order, the only property of the two-body interaction that
enters is the scattering length. These results have been used to
calculate the leading corrections beyond mean-field theory to the
properties of trapped Bose gases\cite{pitaevskii}.  Terms
of order $na^3$ and higher depend on other properties of the interaction, as
discussed in detail recently in Ref. \cite{BN}.

The expansions leading to the above results are asymptotic, and
consequently they give little guidance to the properties of a dense
gas, $na^3\ga 1$.  Even though the range of the interaction is small
compared with the particle separation, correlations between particles
are very important, as one can see from the fact that all terms in the
low-density expansion Eq. (\ref{dilute}) are formally of the same order of
magnitude when $na^3 \simeq 1$.  

The Jastrow wave function \cite{Jastrow}
\bea
   \Psi_J({\bf r}_1,...,{\bf r}_N)= \prod_{i<j}f({\bf r}_i-{\bf r}_j).
\label{jas}
\eea
provides a good approximation for cold dense Bose systems such as the 
hard-sphere gas and liquid $^4$He.  
In these systems the pair correlation 
function $f(r)$ may be determined variationally by minimizing the 
expectation value of the energy,
$E/N = \langle \Psi |H| \Psi \rangle \ /\ \langle \Psi | \Psi \rangle$,
which may be calculated by Monte Carlo methods \cite{HSVMC}. 
The hard-sphere gas problem has also been solved, essentially exactly, 
by quantum Monte Carlo methods \cite{HSQMC} for scattering lengths
almost as large as $r_0$.

A much simpler method, the lowest order constrained variational 
(LOCV) method \cite{BP}, provides a fair approximation to the results of 
quantum Monte Carlo calculations.  The basic idea of this method 
is that for $r < r_0$ the Jastrow function
$f(r)$ approximately obeys the 
Schr\"odinger equation for a pair of particles,
\bea
\left[-\frac{\hbar^2}{m}\frac{d^2}{dr^2} +v(r)\right] rf(r) =\lambda\, rf(r)
   \, , \label{Sch}
\eea
where $\lambda$ is the pair energy, while for $r\gg r_0$, $f$ tends to unity.  
To take into account many-body effects, which become important
when $r$ is comparable to $r_0$,
we impose the condition that
$f(r>d)=1$ and $f^{\prime}(r=d)=0$ at some healing distance $d$.
In the LOCV method the expectation value of the energy is approximated by 
the two-body cluster contribution
\bea 
E/N = 2 \pi n \lambda \int_0^d f^2(r) r^2 dr \ ,
\label{tbe}
\eea
and $d$ is chosen such that 
\bea 
4 \pi n \int_0^d f^2(r) r^2 dr = 1 \ .
\label{int}
\eea
This ensures that on average there are only correlated pairs, 
and $E/N = \lambda /2$.  The energy of a pair is $\lambda$, and thus the calculation 
of the energy by evaluating the expectation value of the Hamiltonian is consistent with the eigenvalue of the 
 Schr\"odinger equation (\ref{Sch}).  

When the range of the interaction is small compared
to both $|a|$ and $d$, as in dilute alkali gases, the interaction 
can be replaced by the boundary condition  
$(rf)'/rf=-1/a$ at $r=0$.  This boundary condition is also satisfied by the  
two-body s-wave scattering wave function, and the scattering 
length $a$ is then the position of the node of $f(r)$ at $\lambda = 0$. 
The solution to Eq. (\ref{Sch}) for positive $\lambda=k^2/m$, is  
$rf(r)\propto\sin[k(r-b)]$, with a node at $r=b$, and $kb=\arctan{ka}$.
In the dilute limit ($ka \rightarrow 0 $), $b$ is equal to $a$, 
and in the dense limit ($ka \rightarrow \infty$), $b=\pi/2k$.  

The other boundary condition, $f^{\prime}(r=d)=0$, gives
\bea 
 \frac{a}{d} = \frac{\kappa^{-1}\tan\kappa -1}{1+\kappa\tan\kappa} 
\,,\label{k}
\eea
where $\kappa =kd$, and the value of $d$ is determined from Eq.~(\ref{int}). 
When $f^2(r)=1$, Eq.~(\ref{int}) gives $d=r_0$. Since $f(r\sim d)\simeq 1$ 
the value of $d$ is generally close to $r_0$. 

In the LOCV approximation $f(r)$ for a short range interaction is 
therefore given by 
\bea
f(r < d) = \frac{d}{r} \frac{\sin[k(r-b)]}{\sin[k(d-b)]} \ .
\label{locvf}
\eea
It is negative for $r < b$, and changes sign at $r = b$.  The Jastrow 
wave function (Eq.~\ref{jas}) does not describe the nodeless ground 
state in this case.  When $a\ll r_0$ the gas has only 
pairwise collisions, and the Jastrow wave function
$\Psi_J$ describes its state accurately. 
We assume that, 
as the gas is compressed or $a$ is increased adiabatically, the nodal 
structure of the wave function is maintained and that $\Psi_J$ 
continues to be a good approximation to the true wave function. 
However, as $a$ or $n$ increases, 
the rate of three-body collisions, which deplete the monoatomic gas 
density by forming molecules, will increase and limit the lifetime of 
the monoatomic gas. 

The variational principle cannot be used to determine the Jastrow
function $f(r)$ in the case of attractive short range interactions,
because the gas is not in its ground state.  In this case, it may
easily be verified that the energy of the gas, as a function of the
healing distance $d$, has an uninteresting minimum at $d \rightarrow
0$ with value zero.  Such a minimum exists for all $\delta$-function
interactions.  However, we anticipate that for the present problem the LOCV method will provide a
good approximation  if we take Eq.\ (\ref{locvf}) for the Jastrow factor. 

In Fig.1 we show the energy per particle for hard-sphereand 
 attractive short range potentials in the LOCV approximation, along
with the low-density expansion (\ref{dilute}) (without the logarithmic
term) and the Lenz expression as a function of $a/r_0$, and the exact
quantum Monte Carlo results for hard-spheres. The LOCV energy
is close to the low density expansion at small values of $a/r_0$, 
though it does not have the correct analytical form.  It contains 
the leading Lenz term in Eq.~(\ref{dilute}), as 
may be verified by expanding the $\tan{\kappa}$ in Eq.~\ref{k} in
powers of $\kappa$ \cite{PS}, but the $(na^3)^{1/2}$ and higher terms 
are approximated. The LOCV energies for the hard-sphere 
gas are in excellent agreement with the exact ones for 
$a/r_0 \le 0.6$, but for larger $a/r_0$ the results for the hard-sphere and 
the attractive short range potentials deviate significantly.

\begin{figure}
\begin{center}
\psfig{file=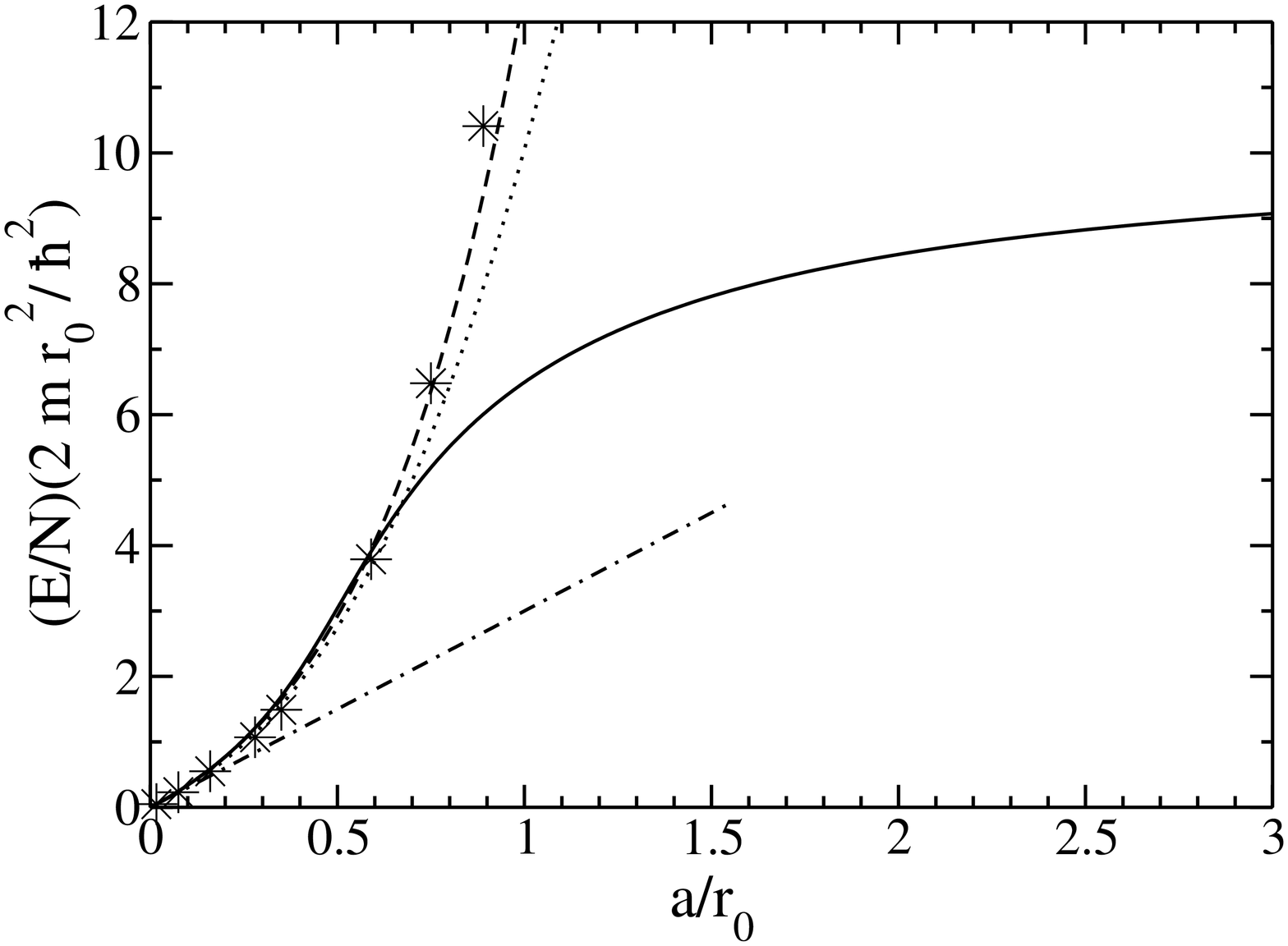,height=9.0cm,angle=-90}
\vspace{.2cm}
\begin{caption}
{Energy per particle, multiplied by $2mr_0^2/\hbar^2$ vs. $a/r_0$. The LOCV 
results for attractive short range and hard-sphere interactions are shown by 
full and dashed lines 
respectively, while the dash-dot and dotted lines show the Lenz result 
and the low density expansion. 
Monte-Carlo numerical results for a hard-sphere gas are shown as stars.} 
\end{caption}
\end{center}
\label{energy}
\end{figure}
  
In the small $a/r_0$ limit, using the result $b=a$, and $d \sim r_0$ 
we find that the LOCV correlation function is given by
\bea
f(r < r_0) = \frac{r_0}{r} \frac{\sin[k(r-a)]}{\sin[k(r_0-a)]} \ , 
\label{dilutef}
\eea
with $k^2 = 3a/r_0^3$.  At small $r$ this correlation is $\sim 1 - a/r$, 
as in the two-body scattering state at zero energy.  

By contrast, when $a/r_0$ is 
large we have $b=\pi/2k$ and Eq. (\ref{k}) reduces to
\bea
  \kappa\tan\kappa = -1 \,,
  \label{kde} 
\eea  
with solutions $kd = \kappa=2.798386.., 6.1212.., ...$, and asymptotically
$\kappa=n\pi$ for integer $n$.  The pair correlation function is given by
\bea
f(r < d) =  \frac{d}{r} \frac{\cos{kr}}{\cos{kd}} \ . 
\label{densef} 
\eea
Substituting this expression in Eq.~(\ref{int}) gives $d=(2/3)^{1/3}r_0$
for $kd=2.798386...$, appropriate for a Jastrow factor $f$ 
with one node.  The energy is given by
\bea
E/N = \frac{\lambda}{2} = \left(\frac{3}{2}\right)^{2/3}
\frac{ \hbar^2(kd)^2}{2m r_0^2} 
    = 13.33 \hbar^2\frac{n^{2/3}}{m}\ . 
\label{hde}
\eea
This energy is proportional to $n^{2/3}$ as one would expect on dimensional 
arguments.  A similar result, $(E/N \propto n^{2/3})$,  
was found for fermions \cite{HH} in the $a \rightarrow - \infty$ 
limit. (We obtain $E/N=15.26\,\hbar^2n^{2/3}/m$ on approximating 
$d$ by $r_0$.)  

For the case of attractive short range interactions it is difficult to
go beyond the LOCV calculation.  When $a/r_0 < 0.2$ we can perform
Monte Carlo calculations of the expectation value of the energy, and
the results are not significantly different from those from the LOCV
method.  At larger values of $a/r_0$ the state described by the
$\Psi_J$ has a clustering instability.  All the atoms end up in a few
dense clusters as the Metropolis walk progresses, as is expected to
happen with time in experiments.

The rate of molecule formation is proportional to the probability of 
finding three atoms together.  In the lowest order (3-body) cluster 
approximation the probability of finding three atoms at points
${\bf r}_1$, ${\bf r}_2$, and ${\bf r}_3$ is proportional 
to $f^2(r_{12})f^2(r_{23})f^2(r_{31})$, where $r_{ij}=|{\bf r}_i-{\bf r}_j|$.
It diverges as all $r_{ij}\to 0$ for the  
$f$ given by Eq.~(\ref{densef}) or Eq.~(\ref{dilutef}). 
However, within the interaction radius 
$R$ the true $f$ does not have the $1/r$ behavior.  We can estimate the 
probability $P_3(R)$ of finding two atoms within the interaction range   
$R$ of a third by approximating the $f$ at $r < R$ by $f(R)$.  In the 
dense limit Eq.~(\ref{densef}) gives $f(r<R) \sim - r_0/R$, and    
\bea
P_3(R) \sim  \left(\frac{r_0}{R}\right)^6 
\left(\frac{4 \pi R^3 n}{3} \right)^2 = 1 \ ,
\label{p3r}
\eea
independent of the gas density.  In contrast, at small values of $a/r_0$ 
Eq.~(\ref{dilutef}) gives $f(r<R) \sim -a/r_0$, and  
\bea
P_3(R) \sim  \left(\frac{a}{R}\right)^6 
\left(\frac{4 \pi R^3n}{3} \right)^2 = \left(\frac{4 \pi~na^3}{3}
\right)^2 \ ,
\label{p3rl}
\eea
At present it is unclear how the behavior can be matched on to the 
$a^4$ dependence predicted for $n|a|^3\ll1$ (see, e.g., Ref. \cite{BBH}).
  
We next estimate the depletion of the condensate.
In the LOCV approximation, the condensate fraction is given by \cite{LOCF} 
\bea
  \frac{n_0}{n} = 1-n\int \left[1-f(r)\right]^2 d^3r \,. \label{CF}
\eea
Substituting Eq.~(\ref{locvf}) in the above equation gives 
\bea
\frac{n_0}{n} = \left(\frac{d}{r_0} \right)^3 \left[ \frac{6}{\kappa^3} 
(\sin \kappa -\kappa \cos \kappa) -1 \right] \ .
\label{locvcf}
\eea
The condensate fraction calculated using the LOCV approximation
is shown in Fig. 2 
along with the predictions for the dilute limit given by  \cite{Beliaev}
\bea
  \frac{n_0}{n} = 1\,-\,\frac{4}{\sqrt{3}\pi} 
\left(\frac{a}{r_0}\right)^{3/2}\ .
 \, \label{CFD}
\eea
Results for the hard-sphere potential obtained with the LOCV method, 
and exact quantum Monte
Carlo calculations \cite{HSQMC} are also shown in Fig. 2.  The LOCV
results for $n_0/n$ are rather crude, since long-range correlations
neglected in the LOCV approximation have a significant effect on
$n_0$.  All the results indicate that superfluidity will be quenched 
when $a\ga r_0$.  In the case of hard-spheres it is known that the 
liquid solidifies at $r_0 \sim a$ \cite{hls71}. 

\begin{figure}
\begin{center}
\psfig{file=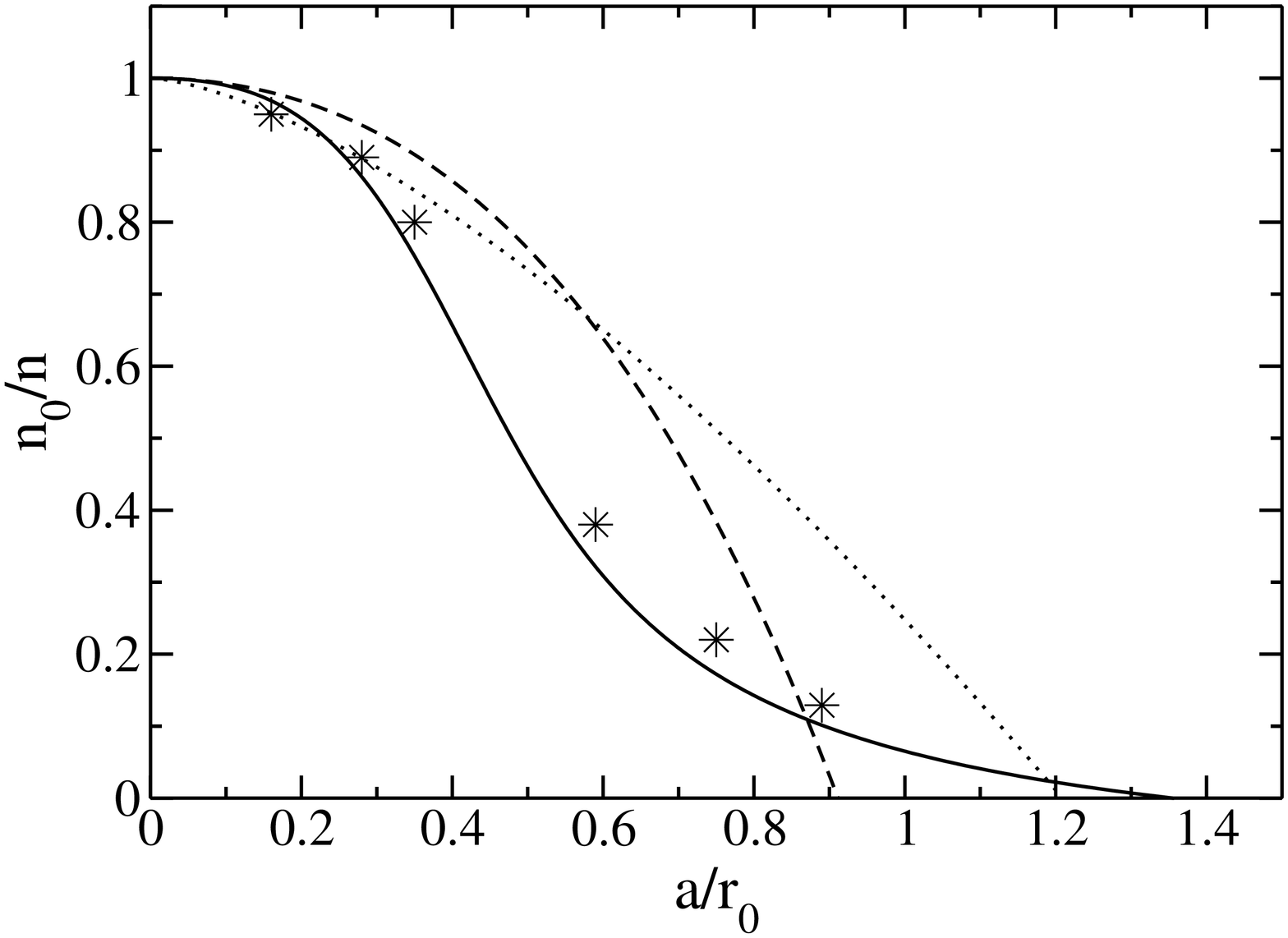,height=9.0cm,angle=-90}
\vspace{.2cm}
\begin{caption}
{Condensate fraction vs. $a/r_0$.  LOCV calculations
for attractive short range
(full line) and hard-sphere (dashed line) interactions, 
Eq.~\ref{CFD} (dotted line) and exact QMC results for hard-spheres (stars). }
\end{caption}
\end{center}
\label{kill}
\end{figure}

The chemical potential at zero temperature is given by $\mu=dE/dN$.
In the dilute limit the leading contributions to $\mu$ are
$\mu= 4\pi\hbar^2na/m\, [1+(32/3)(na^3/\pi)^{1/2}]$,
while, in the dense limit, the LOCV approximation
gives 
\bea
   \mu = 22.22\, \frac{\hbar^2n^{2/3}}{m} \, .  
\eea
The  calculated $\mu(n)$ is shown in Fig. 3. 

The square of the sound speed $c_s$ is given by
$c_s^2 =(n/m) d\mu/d n$, and this 
is also shown in Fig. 3.  Its dilute and dense limits are given by 
$c_s^2=4\pi\hbar^2na/m^2\, [1+16(n a^3/\pi)^{1/2}]$
and 
\bea
   c_s^2=  14.81\, \frac{\hbar^2n^{2/3}}{m^2} \,, 
\eea
respectively. 
At high densities the collective modes in the hydrodynamic limit
are similar to those for fermions 
\cite{Bruun} since the energy per particle scales with
density as $n^{2/3}$, as it does for a free Fermi gas.
The energy per particle in a one-component Fermi gas,
$E/N=(3/10)\hbar^2(6\pi^2n)^{2/3}/m$, is smaller by a factor 2.92 than the
LOCV result in the
dense Bose limit. The chemical potential and sound speed squared
are therefore correspondingly larger in a dense Bose gas.

\begin{figure}
\begin{center}
\psfig{file=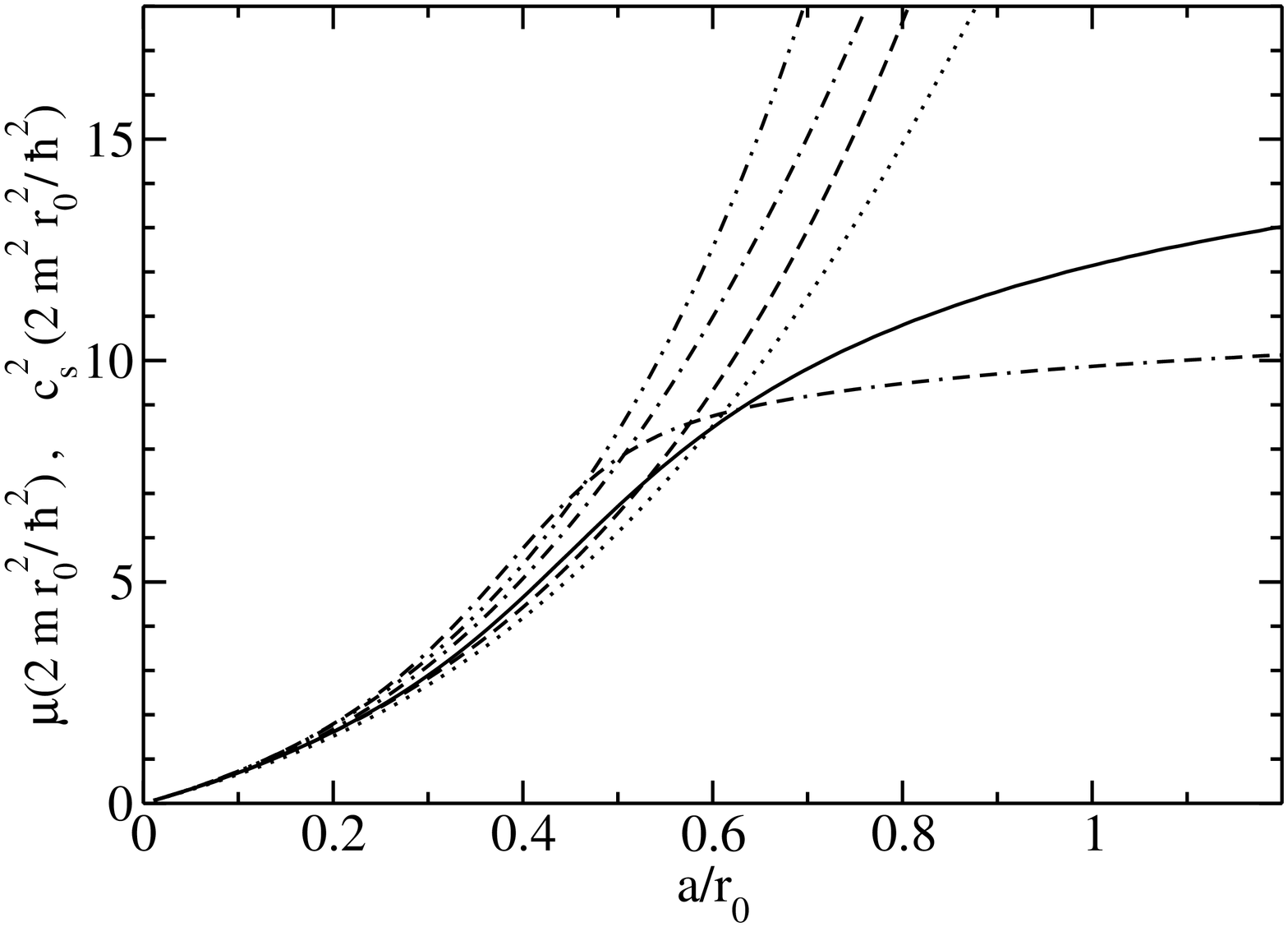,height=9.0cm,angle=-90}
\vspace{.2cm}
\begin{caption}
{ The zero temperature chemical potential multiplied by $2mr_0^2/\hbar^2$
for: attractive short range (solid line) and hard-sphere (dashed line) 
from LOCV calculations,
low density expansion (dotted line) 
and the square of hydrodynamic sound speed multiplied by 
$2m^2r_0^2/\hbar^2$ for:
LOCV for attractive short range (double dash-dot), 
LOCV for hard-sphere(dash-double dot) and low density expansion (dash-dot).} 
\end{caption}
\end{center}
\label{nodes}
\end{figure}

The calculations of the energy density and the estimate of the 
three-body cluster probability $P_3(R)$ indicate that in 
limit where the scattering length is large compared to the
interparticle spacing, the scattering length becomes irrelevant.
These predictions could be investigated experimentally by measuring the
size of a cloud of condensate and the particle lifetime as the
scattering length is varied by exploiting a Feshbach resonance.

We thank G. Baym, S. Jonsell, and J. Yngvason for discussions.
One of us, IEM, acknowledges financial support from 
the program ``Universities of Russia'' (project 015.01.01.04)
and the Russian Ministry of Education grant. no. E00--3--12.
The work of SC, JM, and VRP is supported in part by US National Science 
Foundation via grant PHY 98-00978.

\end{document}